\documentclass[3p,times]{elsarticle} 
 
\usepackage{amssymb}
\usepackage{amsmath} 
\usepackage{amsxtra}
\usepackage{mathrsfs}
\usepackage{bm} 
\usepackage{empheq}
\usepackage{bigints}

\usepackage{subfig}
 
\usepackage{xcolor}

\begin{document}

\begin{frontmatter} 


\title{Mechano-chemical modeling of glia initiated secondary injury of neurons under mechanical load}

\author[add1]{Debabrata Auddya}
\author[add1]{Shiva Rudraraju}
\ead{shiva.rudraraju@wisc.edu} 
\address[add1]{Department of Mechanical Engineering, University of Wisconsin-Madison, WI, USA}

\begin{abstract}
Traumatic Brain Injury (TBI) is associated with an impact or concussion to the head with the injury being specifically characterized through pathological degradation at various biological length scales. To quantify the sequence of events following TBI, various mechanical modeling techniques have been proposed in the literature that seek to quantify neuronal-scale to tissue-scale metrics of brain damage. Broadly, the two categories of degradation encompass physiological deterioration of neurons and upregulation of chemical entities such as neurotransmitters which causes initiation of downstream pathophysiological effects. 
Despite the plethora of pathways which contribute to the downstream processes, in this work, we delineate and model a potential glia-initiated injury pathway that leads to secondary injury. 
The primary aim of this work is to demonstrate a continuum framework which models the multiphysics of mechano-chemical interactions underlying TBI. Using a coupled PDE (partial differential equation) formulation and FEM (finite element method) discretization, the framework highlights evolution of field variables which spatio-temporally resolve mechanical metrics and chemical species across neuronal clusters. The modeling domain encompasses microglia, neurons and the extracellular matrix.
The continuum framework used to model the mechano-chemical interactions assumes a three dimensional viscoelastic network to capture the mechanical response underlying proteins constituting the neuron microstructure and advection-diffusion equations modeling spatio-temporal evolution of chemical species. We use this framework to numerically estimate different critical concentrations of chemical species originating as a result of mechanical strain distribution across the domain.
 In this work, we identified expression of critical biomarkers within the labyrinth of molecular pathways
and construct a mathematical framework which captures the underlying multiphysics of mechano-chemical interactions. This framework is an attempt to quantify secondary injury and thus assist in developing targeted TBI treatments. 
\end{abstract}


\end{frontmatter}


\section{Introduction}
\label{section: introduction}
A broad classification of injuries caused due to an impact or concussion to the head is referred as Traumatic Brain Injury (TBI). The difficulty in TBI diagnoses originates from the heterogeneity of the brain, structural complexity of the neuronal microstructure and the complex interplay of proteins at different length and time scales \cite{kenzie2017concussion}. TBI is generally manifested as primary and secondary, with primary referring to the immediate mechanical response of the brain and the secondary type culminating as a biochemical response \cite{mckee2015neuropathology}. The initiation of the secondary injury is triggered by a pathological grade of mechanical load and as the initial trauma unfolds, a cascade of complex cellular and molecular events follow, potentially exacerbating the damage inflicted during the primary injury. The injury progression involves upregulation of critical neurotransmitters, inflammasomes and prolonged activation of ionic channels which are crucial to maintaining physiological homeostasis within neurons and neuronal clusters \cite{freire2023cellular}. Identifying the cascade of these molecular events in accelerating neurological degradation remains as a fundamental challenge in post injury diagnoses. In this work we aim to qualitatively and quantitatively demonstrate one of the many pathways initiated by glial cells , underlying TBI secondary injury. \par 
Previous studies have highlighted key biomarkers relevant to TBI which trigger downstream pathological pathways
\cite{ghaith2022literature}. Recent studies have highlighted activation of the transmembrane channel proteins Pannexin-1 (Panx1) \cite{seo2021pannexin} present in neurons and glial cells, due to an increase in mechanical strain \cite{albalawi2017p2x7}, which are key mediators of ATP (Adenosine Triphosphate) release into the extracellular region, thereby initiating neurodegeneration \cite{ebanks2022mitochondrial}. An increase in extracellular ATP, which is widely known for its role in energy metabolism, activates trimeric purinergic receptors P2X which are ATP-sensitive ion channels present in the microglia \cite{hattori2012molecular}. Xing et al. \cite{xing2016modeling} quantitatively highlighted the influence of increasing extracellular ATP concentration on P2X and P2Y receptor sensitivity and activation dynamics of the P2X7 receptor, whose expression was associated with inflammation \cite{andrejew2020p2x7, rotondo2022role}. 
Further, the P2X7 receptor also assists in production and release of TNF-$\alpha$ (Tumor Necrosis Factor), another major pro-inflammatory cytokine, responsible for triggering an inflammatory response \cite{you2021tumor,barbera2017p2x7}.  
TNF-$\alpha$ has been implicated in modulating glutamate transmission and excitotoxicity \cite{nicosia2024glutamate} in the brain and can increase the release of glutamate, promoting excitotoxicity by several mechanisms \cite{olmos2014tumor} 
such as enhancing expression of the NMDA receptor \cite{jara2007tumor} and disrupting glutamate clearance mechanisms \cite{o2017role} in the brain which leads to neuronal damage \cite{takeuchi2006tumor}. Together they cause irregular signal transmission across neurons and uncontrolled inflammation leading to neurodegeneration \cite{jia2015taming,vaglio2024pathological}. \par
To address this complexity in spatio-temporally quantifying downstream neurodegenerative processes, we identify a clinically relevant ensemble of biomarkers constituting glia-induced secondary injury. 
Additionally, we demonstrate spatial localisation of these species and their downstream influence on associated elements within the pathway. Therefore, our objective is to quantify the gap between isolated molecular events at the cellular scale and neurodegeneration at the tissue scale using a qualitative narrative. \par
Our approach of quantifying the mechanotransduction process underlying TBI is structured into two components. Firstly, we construct a mechano-chemical formulation consisting of relevant chemical and mechanical fields which dictate the multi-physics of secondary injury. Secondly, we numerically develop geometrical domains resembling neurons, microglia and extracellular matrix (ECM) to spatially represent localisation and evolution of chemical species within a larger neuronal cluster domain. This enables visualisation of the interactions between mechanical and chemical fields and their spatio-temporal resolution within single and multiple neuron-microglia-ECM assembly. 
Underlying this multiphysics setup for approximating mechanics induced chemical injury, the mechanical response is modeled using a three dimensional viscoelastic network consisting of proteins which confer structural stability to the neuron-microglia-ECM assembly. The chemical response is modeled using advection-diffusion equations with chemical fields representing molecular species (ATP, TNF-$\alpha$, P2X7, glutamate) and geometrical localisation accounting for the spatial heterogeneity of these constituents across the domain. The finite element (FE) method is used to discretize the underlying coupled PDE formulation. Using appropriate sets of boundary conditions across the domain, we demonstrate spatio-temporal evolution of biomarkers and characterize mechano-chemical basis of injury thresholds. \par 
To summarize we propose a multiphysics model of molecular mechanotransduction and capture the mechano-chemical dynamics arising out of secondary injury. In Section \ref{section:necroptosis} we provide a detailed description of the chemical pathway proposed. In Section \ref{section:mathematical} we introduce the biomarkers and quantify them using PDEs. This is followed by the numerical formulation of the multiphysics setup. In Section \ref{section:results} we discuss about specific boundary value problems (BVPs) similar to TBI conditions and demonstrate mechano-chemical interactions. Additionally we construct a computational injury curve and increase the phase spectrum for TBI diagnoses. We conclude with Section \ref{section:conclusion} with a brief discussion and directions for future possibilities in the computational treatment of TBI. 

\section{Chemical Pathway of Necroptosis}
\label{section:necroptosis}
TBI is often addressed as a "biphasic injury" having a primary and a secondary component \cite{ng2019traumatic}. While experimental and clinical studies have established treatment strategies for primary injury, a detailed understanding of secondary injury remains unclear 
primarily due to the existence of the plethora of degradation pathways \cite{maas2008moderate}. 
This study aims to elucidate one such mechanism driving secondary injury, focusing on identifying,  quantifying biomarkers, and spatio-temporally modeling the proposed pathway that leads to necroptosis \cite{hu2022role}.
\par
Cell death, traditionally categorized into three primary types namely apoptosis, autophagy, and necrosis showcases distinct morphological and biochemical transformations, each exhibiting unique pathways during cellular degradation \cite{healy1998apoptosis}. In the context of TBI, cellular damage is characterized through apoptosis and necrosis \cite{fink2005apoptosis}. 
In the context of secondary injury, emerging research suggests the presence of a meticulously regulated form of cell death, termed as necroptosis \cite{nie2022role,hu2022role} which is characterized by controlled activation and programmed cell death unlike apoptosis \cite{akamatsu2020cell} or necrosis \cite{dhuriya2018necroptosis}. 
In this communication, our objective is to intricately delineate and quantify a possible chemical pathway governing necroptosis. 
\par
\begin{figure}[h!]
    \centering
    \includegraphics[width=0.8\linewidth]{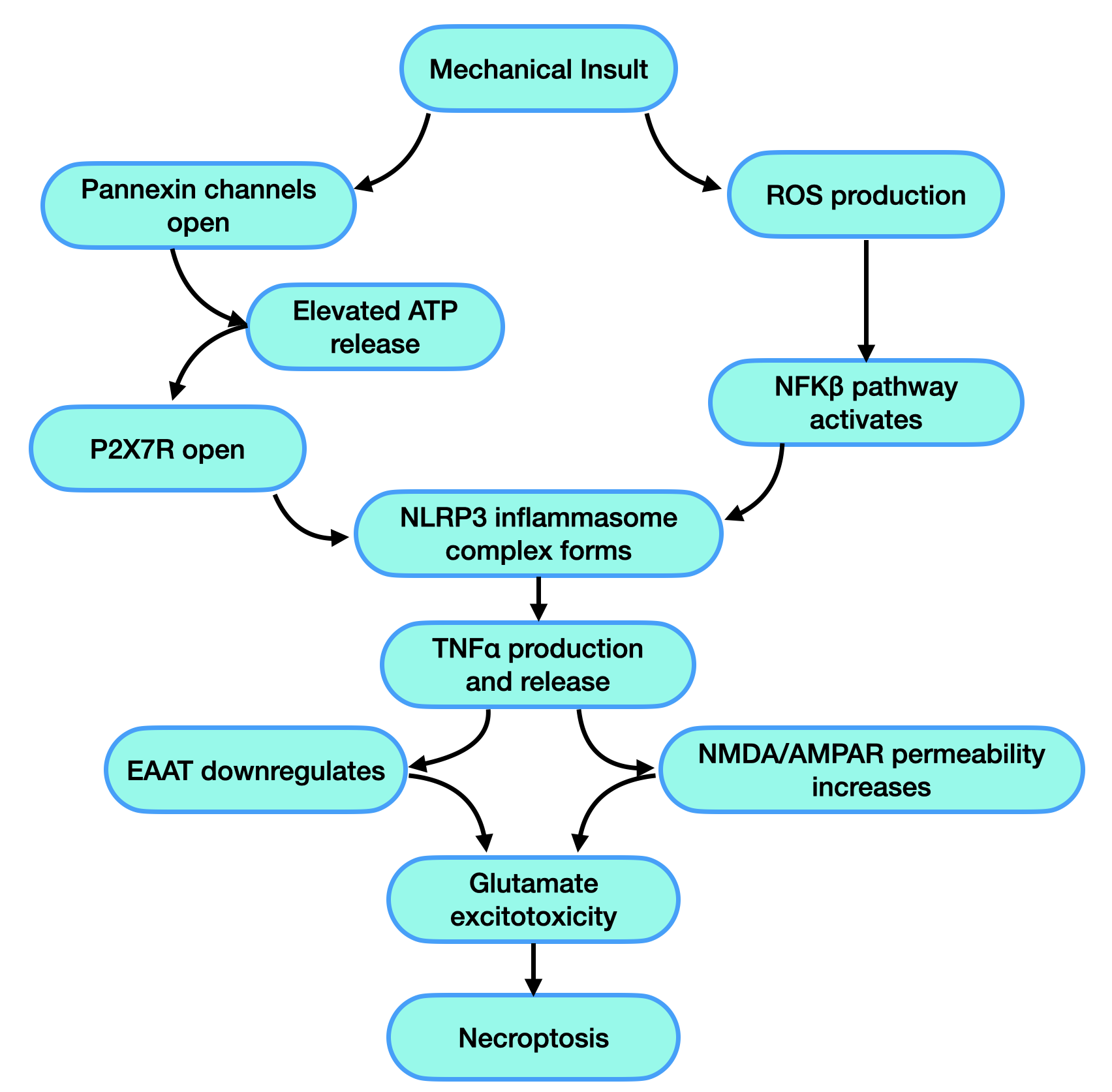}
    \caption{Proposed pathway of necroptosis underlying secondary injury during TBI. The pathway begins with mechanical deformation manifested as increased strain to the brain, which triggers opening of the pannexin channels leading to massive ATP efflux into the extracellular region. Increased ATP causes purinergic receptors to activate, particularly the P2X7R, which causes potassium efflux from the intra- to extracellular milieu. Decrease in ionic concentration of potassium initiates formation of the NLRP3 inflammasome complex. Simultaneously there is a heightened increase in the demand for oxygen causing oxidative stress and production of reactive oxygen species. This causes the NF$\kappa$B pathway to trigger, also leading to intracellular formation of the NLRP3 inflammasome complex. This complex is responsible for modulating formation of pro-inflammatory products mainly cytokines such as TNF-$\alpha$ and ILs. TNF-$\alpha$ causes dysregulation in neuronal signalling by blocking EAAT's on astrocytes thereby reducing glutamate uptake. It also intensifies excitatory transmission by increasing permeability of the glutamate receptors to calcium ions. Increased glutamate concentration in the post synaptic region causes excitotoxicity and is considered as the pivotal step towards necroptosis.}
    \label{fig:pathway}
\end{figure}

Following an insult, a cascade of downstream molecular pathways initiate within the brain and here we refer them as biomarkers.
Crucial to biomarker expression is the activation of microglia, the resident immune cells of the central nervous system which causes production and release of molecules such as NLRP3 inflammasome \cite{o2020nlrp3}, pro-inflammatory cytokines such as interleukins \cite{yan1996rage}, TNF-$\alpha$ \cite{woodcock2013role, longhi2013tumor} and neurotransmitters such as glutamate \cite{guerriero2015glutamate} most of which are expressed in varying degrees during a concussion as illustrated in Fig.\ref{fig:pathway}.
The pathway begins with a concussion to the brain characterized by mechanical deformation within neurons and neuron clusters and terminates with glutamate excitotoxicity, a classic indicator suggestive of neuronal degeneration \cite{armada2020going}. 
Initiating the molecular cascade upon concussion, mechanosensitive receptor channels in the microglia, identified as Pannexins, become activated as the first elemental response in the pathway \cite{bao2004pannexin}. 
Recent studies have demonstrated the influence of mechanical deformation in prolonged activation of pannexin channels which trigger excessive release of ATP \cite{albalawi2017p2x7,xia2012neurons}. \par
Excess extracellular ATP, beyond its normal physiological levels, can trigger several detrimental effects including neuronal excitotoxicity \cite{choi2021extracellular}, inflammation \cite{cauwels2014extracellular}, oxidative stress \cite{cruz2007atp} and impaired synaptic function \cite{vroman2014extracellular}. This surplus ATP present in the extracellular space can overstimulate purinergic receptors which 
are widely expressed in microglia \cite{calovi2019microglia}. Of particular interest is the P2X7 receptor, a distinct member of the purinergic receptor family of ligand-gated ion channels which plays multifaceted roles in various physiological and pathological processes \cite{andrejew2020p2x7}. Xing et al. \cite{xing2016modeling} characterized P2X receptor responses across varying concentrations of ATP, examining both human and rodent receptors. Activation of these receptors, typically at high concentrations of ATP allows the influx of calcium and sodium ions and subsequent efflux of potassium ions \cite{piccini2008atp, xu2020distinct}. 
which causes hyperexcitability \cite{florence2012extracellular}, altered synaptic activity \cite{tagluk2014influence} and neuronal degradation \cite{munoz2013k+,koumangoye2022role}. A reduction in intracellular potassium concentration
serves as a signal for NLRP3 inflammasome \cite{munoz2013k+} activation in microglia, 
and consequently leads to the processing and release of pro-inflammatory cytokines into the extracellular region \cite{blevins2022nlrp3}. In addition to the molecular pathway under consideration, an increased post injury oxygen consumption in the brain leads to heightened production of reactive oxygen species (ROS) \cite{fesharaki2022oxidative} which directly stimulates the production of TNF-$\alpha$ \cite{liang2017reactive}.
\par
TNF-$\alpha$ is a pleiotropic pro-inflammatory cytokine which is primarily released by activated microglia and other immune cells in the brain during secondary injury \cite{chio2015microglial}. 
Despite its beneficial in immune defence and tissue repair \cite{shohami1999dual,francisco2015tnf}, in this study we focus on the pathophysiological impacts of this cytokine and its influence on downstream molecules. 
TNF-$\alpha$'s detrimental impact within the molecular pathway manifests notably through its influence on glutamate excitotoxicity \cite{takeuchi2006tumor}. Glutamate, 
often revered as the "commander" of the brain \cite{mattson2023sculptor} orchestrates several pivotal functions crucial for neuronal signalling \cite{reiner2018glutamatergic}, synaptic plasticity \cite{nadler2012plasticity}, memory and cognition \cite{pal2021glutamate}. 
Neurons initiate the process of glutamate synthesis by transforming glutamine into glutamate and subsequently releasing it to astrocytes which capture the released amino acid, converting it back into glutamine before transporting it back to neurons. During pathophysiological conditions, TNF-$\alpha$ downregulates the expression and function of glutamate transporters on astrocytes, such as excitatory amino acid transporters (EAATs). The interference of the inflammatory cytokine with glutamate uptake by astrocytes leads to increased extracellular glutamate levels, contributing to excitotoxicity and neuronal damage \cite{olmos2014tumor, guerriero2015glutamate}. \par
Within glial cells, the activation of tumor necrosis factor receptor 1 (TNFR1) by TNF-$\alpha$ elicits a dual effect. Firstly it stimulates the process of glutamate removal from astrocytes and secondly it impedes their ability to efficiently clear glutamate from the synaptic cleft. 
Simultaneously, TNF-$\alpha$ decreases the surface presence of inhibitory gamma aminobutyric acid (GABA)-A receptors, further tipping the balance towards increased glutamate \cite{pribiag2013tnf}. 
The consequences of glutamate excitotoxicity commences at the molecular level, where dysregulated glutamate signaling disrupts neuronal homeostasis, triggers necroptosis \cite{arrazola2019compartmentalized,li2008necroptosis} and also contributes to chronic neurodegeneration, culminating in conditions such as Alzheimer's disease \cite{wang2017role}, Parkinson's disease \cite{iovino2020glutamate} and Huntington's disease \cite{andre2010dopamine}. 
Incorporating the qualitative insights from the proposed pathway we have developed a PDE based multiphysics (mechano-chemical) framework. We use our framework primarily to model neurons, neuron clusters and their microenvironment to visualise and resolve the spatio-temporal dynamics of the molecular biomarkers. 
\subsection{Reduced pathway considered for quantification of chemical species}
\begin{figure}[h!]
    \centering
    \includegraphics[width=0.8\linewidth]{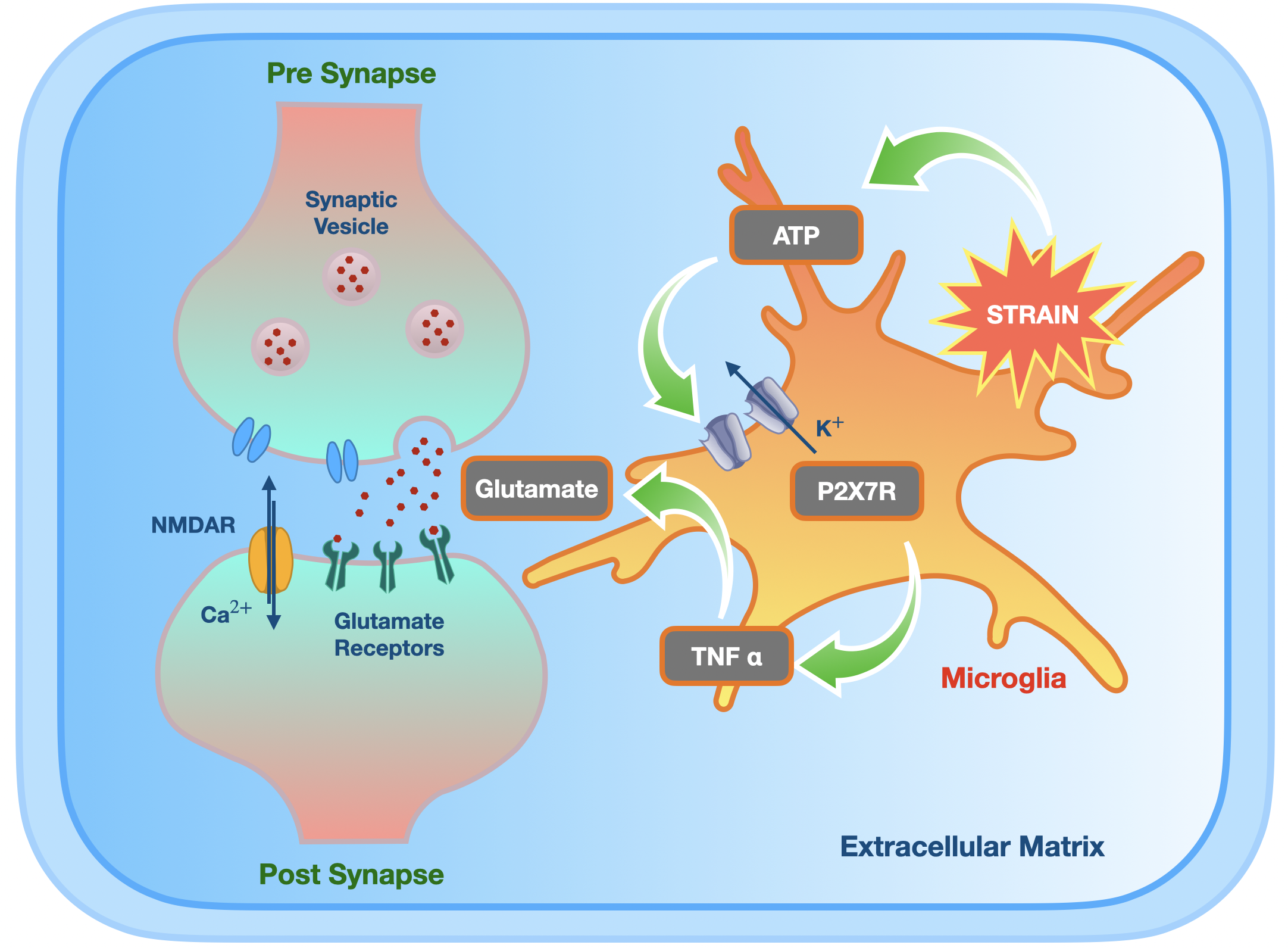}
    \caption{Representation of a reduced pathway consisting of quantifiable elements which contribute towards necroptosis. Illustrated in this pathway are ATP, TNF-$\alpha$ and glutamate classified as chemical concentration fields and the P2X7 receptor mathematically as a phase field. This is attributed qualitatively to the opening probability of the channel upon prolonged activation by ATP. The shortened pathway enables formulation of diffusive equations to spatio-temporally resolve the chemical species across the inhomogenous neuronal landscape.}
   \label{fig:shortpathway}
\end{figure}

Given the challenges and complexities in mathematically modeling the entire pathway in Fig.\ref{fig:pathway}, primarily due to insufficient quantifiable data for every element, we suggest a more succinct approach. Our proposal involves a reduced pathway consisting of fundamental chemical 
biomarkers within the cascade, supported by substantial experimental validation found in the literature Fig.\ref{fig:shortpathway}. \par
As illustrated in Fig.\ref{fig:shortpathway} the cascade initiates with mechanical deformation, manifested as strain within neurons and neuron clusters. Mathematically, the imposed boundary condition is treated as deformation while the corresponding strain response is obtained from the viscoelastic constitutive network of the underlying neuronal microstructure. Using this strain field generated across the domain of interest (neurons or neuron clusters), we use it as an input for driving ATP generation. This assumption is hypothesized from the phenomena that mechanical deformation causes activation of Panx1 channels leading to ATP release from the microglia to the extracellular region. \par
Excess concentration of extracellular ATP causes activation of the purinergic P2X7 receptors situated in the microglia which facilitate formation of pro-inflammatory complexes. Recent findings \cite{xing2016modeling} have quantified the effect of ATP on members of the P2X and P2Y receptor family which correlate the effect of ATP concentration with P2X7 receptor opening probability. In the modeling process, the purinergic receptor channel has been estimated numerically as a phase field. The consequences of this activation leads to production of enzymes, formation of inflammasomes and release of cytokines from the microglia such as TNF-$\alpha$. \par
The release of TNF-$\alpha$ is regulated heavily by modulation of the P2X7 receptor. Studies have indicated increase in concentration of TNF-$\alpha$ due to receptor activation. This inflammatory molecule is mathematically modeled as a diffusive chemical field and driven by a function which depends on the P2X7 phase field. While there are a number of crucial elements preceding the formation and release of TNF-$\alpha$, they have not been considered in the reduced pathway due to a paucity in quantifiable data relating intermediate molecules with any of the cascade components considered. Due to the increased extracellular concentration of TNF-$\alpha$, the microglial uptake of glutamate reduces in the glial cells and subsequently increases in the post synaptic region. In our reduced pathway, glutamate is also modeled as a diffusive chemical field localised in the neuronal synapses. Our modeling process is facilitated by data associating TNF$-\alpha$ with glutamate, where evolution of the latter is driven by a function relating the former. \par 
We discuss the multiphysics of the functional dependencies of all of the above chemical species' and derive reaction terms which demonstrate quantifiable relations between them. 
The idea is to present a novel mathematical framework incorporating a first-of-its-kind necroptotic pathway modeled with a rigorous numerical implementation using state-of-the-art scientific computing tools in FEM and demonstrate mechanical and chemical metrics underlying secondary injury. The computational framework is made available to the wider research community as an open source library \cite{GitRepo2024}. 

\section{Mathematical Formulation}
\label{section:mathematical}
The first step in the multiphysics formulation is identifying the biological species constituting the molecular mechanotransduction pathway and characterizing their diffusive dynamics, sources and chemo-mechanical interactions. We consider concentrations of the following chemical species as the primary fields: ATP ($c_a$), TNF-$\alpha$ ($c_t$), Glutamate ($c_g$) and ionic channels represented as a phase field: P2X7 ($\phi_p$). The idea behind a phase field assumption of ionic channels is the appoximation of gating probability (0-closed, 1-open) due to concentration sensitivity of certain chemical species. The evolution of these chemical fields are modeled using the following advection-diffusion equations:
\begin{gather} 
    \frac{\partial c_a(\textbf{x},t)}{\partial t} = \nabla \cdot (D_a \nabla c_a) +f(\epsilon,\dot{\epsilon}), \hspace{0.5cm} \textbf{x} \in \Omega \label{eq:atp} \\
    \frac{\partial c_t(\textbf{x},t)}{\partial t} = \nabla \cdot (D_t \nabla c_t) +f(c_a, \phi_p), \hspace{0.5cm} \textbf{x} \in \Omega \label{eq:tnf}\\
    \frac{\partial c_g(\textbf{x},t)}{\partial t} = \nabla \cdot (D_g \nabla c_g) +f(c_t), \hspace{0.5cm} \textbf{x} \in \Omega \label{eq:glu}\\
    \phi_p = \Phi(c_a) \label{eq:p2x7}
\end{gather}
where the source terms are expressed as $f(\epsilon, c_a, c_t, \dots)$ and $\Phi$ represents receptor sensitivity as a function of concentration of ATP. The diffusivity terms are expressed as $D_x$ where the subscript $x$ refers to individual species. \par
\begin{figure}[ht]
    \centering
    \includegraphics[width=0.7\linewidth]{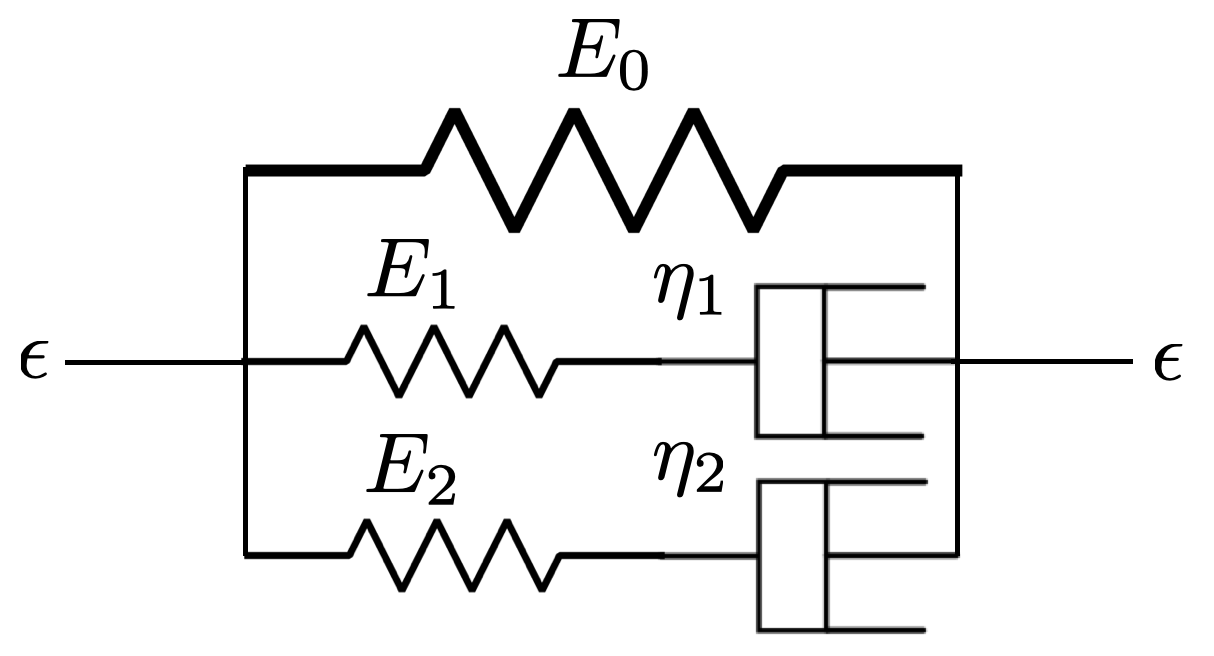}
    \caption{Mechanical network representing the neuronal microenvironment. The mechanical estimates of springs and dampeners are obtained from Wang et al. \cite{wang2022viscoelastic}}
    \label{fig:mechNetwork}
\end{figure}

In order to model the mechanical response of the underlying neuronal microstructure, a viscoelastic network consisting of three dimensional springs and dampeners have been considered. The network illustrated in Fig.\ref{fig:mechNetwork}, capable of mimicing the extracellular matrix is similar to the work of Wang et al.\cite{wang2022viscoelastic} who have characterised the mechanical properties of the elements constituting the network. The response of this structure is modeled using the conservation of linear momentum. 
\begin{align}
    \nabla \sigma = 0
\end{align}
\subsection*{Variational Formulation}
Casting the above governing equations in their variational
(integral/weak) form. This formulation is used to solve these equations within a standard finite element method framework. We consider treatment of the chemical species dynamics followed by the mechanical equilibrium equations. \par
Find the primal concentration fields \{$c_a, c_t, c_g, \phi_p, \textbf{u}$\} where,
\begin{align}
c_i \in \mathscr{S}_{c_i}, \hspace{0.5cm} \mathscr{S}_{c_i} = \{c_i \in \mathscr{H}^1(\Omega) \hspace{0.3cm}|c_i = \Bar{c_i} \hspace{0.3cm} \forall \hspace{0.3cm} \textbf{X} \in \Gamma^{c_i}_{g} \}
\end{align}
where $i \in \{ \text{ATP}, \text{TNF-}\alpha, \text{Glutamate} \} $ and the phase field, 
\begin{align} 
\phi_p \in \mathscr{S}_{\phi_p}, \hspace{0.5cm} \mathscr{S}_{\phi_p} = \{\phi_p \in \mathscr{H}^1(\Omega) \hspace{0.3cm}|\phi_p = \Bar{\phi_p} \hspace{0.3cm} \forall \hspace{0.3cm} \textbf{X} \in \Gamma^{\phi_p}_{g} \}
\end{align}
finally displacement, 
\begin{align}
    \textbf{u} \in \mathscr{S}_{\textbf{u}}, \hspace{0.5cm} \mathscr{S}_{\textbf{u}} = \{\textbf{u} \in \mathscr{H}^1(\Omega) \hspace{0.3cm}| \textbf{u} = \Bar{\textbf{u}} \hspace{0.3cm} \forall \hspace{0.3cm} \textbf{X} \in \Gamma^{\textbf{u}}_{g} \}
\end{align}
such that for all variations,
\begin{gather}
w_{c_i} \in \mathscr{V}_{c_i}, \hspace{0.5cm} \mathscr{V}_{c_i} = \{w_{c_i} \in \mathscr{H}^1(\Omega) \hspace{0.3cm}|w_{c_i} = 0 \hspace{0.3cm} \forall \hspace{0.3cm} \textbf{X} \in \Gamma^{c_i}_{g} \} \\
w_{\phi_p} \in \mathscr{V}_{\phi_p}, \hspace{0.5cm} \mathscr{V}_{\phi_p} = \{w_{\phi_p} \in \mathscr{H}^1(\Omega) \hspace{0.3cm}|w_{\phi_p} = 0 \hspace{0.3cm} \forall \hspace{0.3cm} \textbf{X} \in \Gamma^{\phi_p}_{g} \} \\
w_{\textbf{u}} \in \mathscr{V}_{\textbf{u}}, \hspace{0.5cm} \mathscr{V}_{\textbf{u}} = \{w_{\textbf{u}} \in \mathscr{H}^1(\Omega) \hspace{0.3cm}|w_{\textbf{u}} = 0 \hspace{0.3cm} \forall \hspace{0.3cm} \textbf{X} \in \Gamma^{\textbf{u}}_{g} \}
\end{gather}
we have, 
\begin{gather} \label{eq:weak}
\bigintsss_{\Omega} w_{\textbf{u}} \sigma dV - \bigintsss_{ \Gamma^{\textbf{u}}_{h}}
     w_u \textbf{t} dS =0 \\
    \bigintsss_{\Omega} w_{c_i} \frac{\partial c_i}{\partial t} dV + \bigintsss_{\Omega} ( D_{c_i} \nabla w_{c_i} \nabla c_i  - w_{c_i} f(c_j,\epsilon) ) dV - \bigintsss_{\Gamma^{c_i}_{h}} w_{c_i} (\nabla c_i \cdot \textbf{n}) dS = 0 \\
    \bigintsss_{\Omega} w_{\phi_p} (\Phi(c_a) - 1.0)dV = 0
\end{gather}
where $w_{c_i}, w_{\phi_p}$ are the variations in chemical concentrations and phase field respectively, $w_{\textbf{u}}$ is the variation in displacement and $i \neq j$. $\Omega$ is the problem domain,$ \{\Gamma^{c_i}_{g}, \Gamma^{\phi_p}_{g}, \Gamma^{\textbf{u}}_{g} \}$ are the Dirichlet boundaries for the chemical concentration fields, phase field and displacement vector respectively. Similarly, $\Gamma^{c_i}_{h}$ and $\Gamma^{\textbf{u}}_{h}$ are the Newmann boundaries for the chemical fields and displacment respectively and $\textbf{n}$ is the unit normal vector. In this formulation, we assume that there are traction boundary conditions $\textbf{t}$ for the displacement boundaries and no chemical species flux at all the boundaries $( \nabla c_i \cdot \textbf{n} = 0)$. \par
The mechanical and chemical properties which have been used from literature are summarized in Table \ref{tab:tableOne}. 
\begin{table}[hbt!]
\caption{Mechanical and Chemical Properties}
\label{tab:tableOne}
\centering
\begin{tabular}{|c|c|c|}
\hline
\textbf{Property} & \textbf{Value} & \textbf{References} \\ \hline  
      $E_0$   &   3 $\mu$N/$\mu$$m^2$    & \cite{wang2022viscoelastic}          \\ \hline  
      $E_1$   &   1 $\mu$N/$\mu$$m^2$    &          \cite{wang2022viscoelastic}  \\ \hline  
     $E_2$   &   130 $\mu$N/$\mu$$m^2$    &         \cite{wang2022viscoelastic}   \\ \hline  
    $\tau_1$    &  16  s   &    \cite{wang2022viscoelastic}        \\ \hline  
    $\tau_2$   &   400 s   &   \cite{wang2022viscoelastic}         \\ \hline  
     bulk &     1000 $\mu$N/$\mu$$m^2$ &          \cite{konno2021modelling} \\ \hline  
     $D_a$    &  300 $\mu$$m^2$/s  &\cite{bennett1995quantal},\cite{zhang2018atp}             \\ \hline  
     $ D_t$   &  150 $\mu$$m^2$/s  & \cite{ross2018diffusion}, \cite{goodhill1997diffusion}           \\ \hline
      $D_g$   &   460 $\mu$$m^2$/s    &    \cite{rusakov2011shaping},\cite{moussawi2011extracellular}   \\ \hline  
\end{tabular}
\end{table}
\subsection*{Multiphysics of Reaction Terms}
The equations representing evolution of the chemical species are driven by source terms which depend on evolution characteristics of other chemical fields. In our formulation, we have introduced four different chemical species and a displacement variable, which are related to each other sequentially. Based on the quantitative evidence from recent literature, the functional dependence of each primary field with others have been established.
\subsubsection*{ATP}
As discussed earlier, when an injury manifests, it does so by activating pannexin channels embedded in the microglia which causes excess energy molecules, ATP, to efflux. While we have established a qualitative relationship between external mechanical deformation and spatial modulation of ATP concentration, there is a lack of quantitative evidence on specific metrics which relates the two phenomena. Therefore, we propose a linear relationship between the amount of mechanical strain and strain rate obtained from the underlying constitutive framework and the amount of ATP produced subsequently. The mathematical relation between the source term in Eq.\ref{eq:atp} driving ATP evolution and mechanical deformation is expressed as:
\begin{align}
f(\epsilon, \Dot{\epsilon}) = K_1 f(\epsilon) + K_2 f(\Dot{\epsilon})
\end{align}
where $K_1$ and $K_2$ are fitting constants with units M and M-sec respectively. The functions $f(\epsilon)$ and $f(\Dot{\epsilon})$ are plotted in Fig.\ref{fig:ATPrxn} below. The evolution profile is presented in Fig.\ref{fig:allFigures}(A). 
\begin{figure}[h!]
    \centering
    \includegraphics[width=\linewidth]{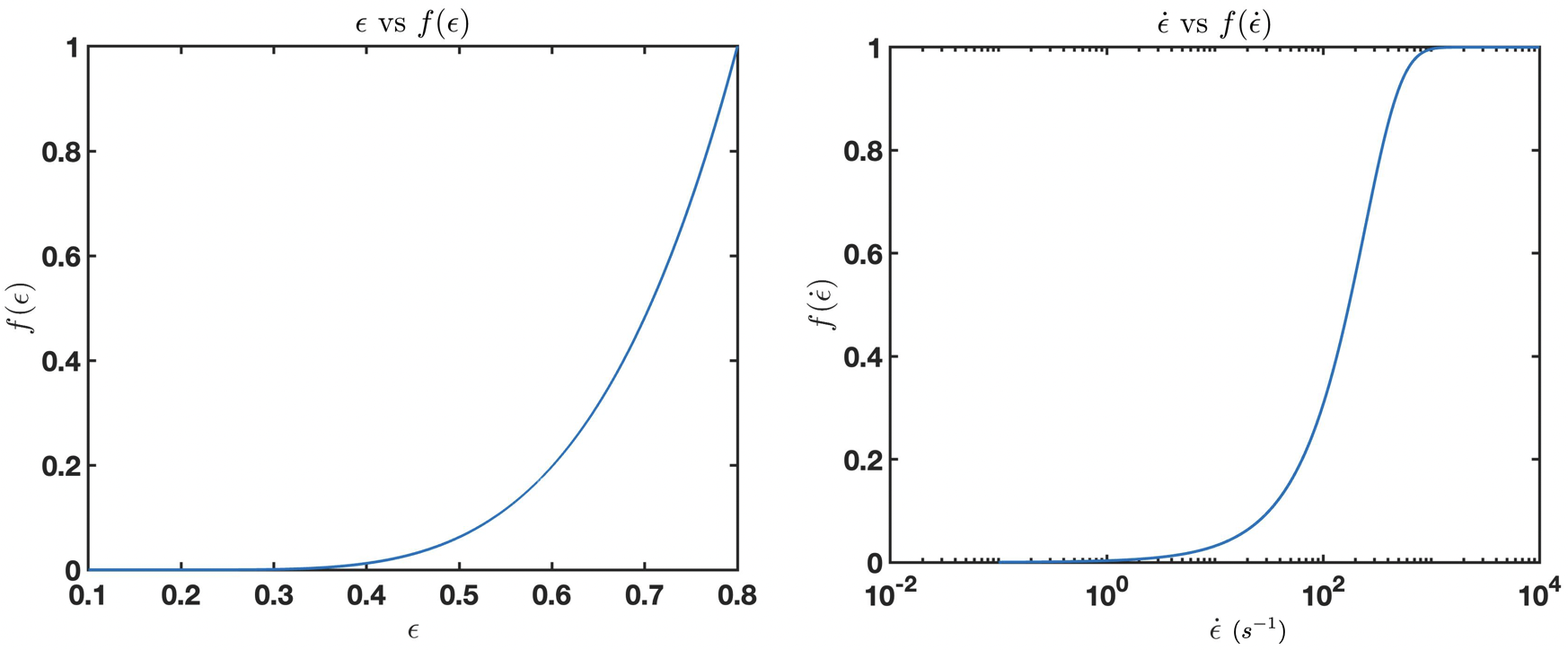}
    \caption{(Left) Dependence of $f(\epsilon)$ with $\epsilon$. (Right) Dependence of $f(\Dot{\epsilon})$ with $\Dot{\epsilon}$. }
    \label{fig:ATPrxn}
\end{figure}
\subsubsection*{P2X7 Receptor}
Xing et al.\cite{xing2016modeling} demonstrated a critical concentration of ATP beyond which the purinergic P2X7 receptor activates leading to increased probability of inflammasome complex formation. Based on the quantitative data provided in their experiments, a tan hyperbolic function has been used to recreate the normalised response of the receptor. We have illustrated (Fig.\ref{fig:allFigures}(B)) and compared the data with our proposed function as a mathematical source term for P2X7 receptor channel properties. 
The specific expression used in Eq.\ref{eq:p2x7} to quantify the phase field evolution representing the purinergic receptor is given as: 
\begin{align}
\Phi_p = \text{tanh} (Af(c_a))
\end{align}
where $A = 10^{2.25} \text{M}^{-1}$ and $f(c_a)$ varies between $10^{-8} - 10^{-1}$ M, which is considered keeping in view the typical ballpark of ATP produced during pathological conditions. 
\subsubsection*{TNF-$\alpha$}
Upon P2X7 receptor activation, a cascade of events unfolds, of which the pivotal step involves release of pro-inflammatory molecules namely TNF-$\alpha$. Based on the work of Barberà-Cremades et al. \cite{barbera2017p2x7}, a relationship between amount of this cytokine release upon P2X7 receptor activation over time has been shown. We compare the experimental findings with another tanh function tailored to incorporate contribution of the purinergic receptor as illustrated in Fig.\ref{fig:allFigures}(B). In this comparison, TNF-$\alpha$ accumulates as a function of time and saturates at a certain concentration. We propose a mathematical model which formulates a source term, as mentioned in Eq.\ref{eq:tnf} by associating the P2X7 receptor based on the experimental findings and spatio-temporally resolving the maximum limits of this chemical species.  
\begin{align}
f(c_a, \phi_p) = B \phi_p
\end{align}
where $B = 0.1$ and a critical concentration of ATP ($c_a$) is considered as a trigger for TNF-$\alpha$ evolution. The reaction profile is shown in Fig.\ref{fig:allFigures}(C). 
\subsubsection*{Glutamate}
\begin{figure}[h!]
    \centering
    \includegraphics[width=0.8\linewidth]{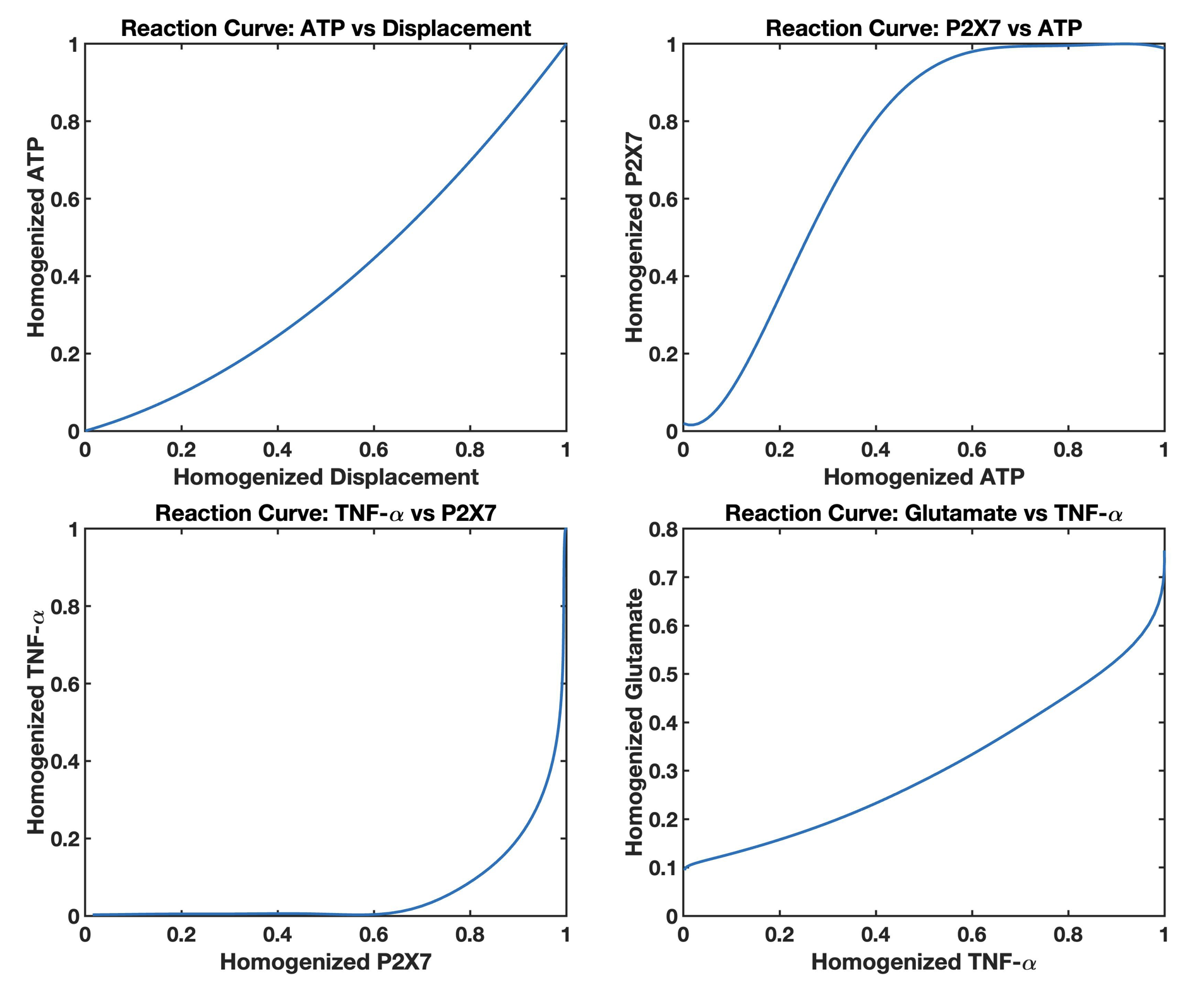}
    \caption{\textbf{(A)} ATP evolution against increasing strain (upto 80 percent) has been shown here with different strain rates. In particular, the strain rate $10^{2} s^{-1}$ has been chosen to further illustrate correlations between the chemical fields. \textbf{(B)} P2X7 receptor opening probability is plotted against ATP evolution for the aforementioned strain rate. Furthermore, this data suggests a positive agreement with the findings of Xing et al. \cite{xing2016modeling} for estimating P2X7 receptor probability with change in ATP. \textbf{(C)} TNF-$\alpha$ evolution with change in receptor opening characteristics has been demonstrated in this plot. It is observed that at a relatively higher value of P2X7, TNF-$\alpha$ diffuses out and continues evolving with time. \textbf{(D)} Glutamate excitotoxicity in the extracellular region, influenced by increased presence of inflammatory cytokines like TNF-$\alpha$ can be observed in this plot. These results are compared with the findings of Zou et al. \cite{zou2005tnfalpha} and show significant resemblance in the nature of glutamate evolution.}
    \label{fig:allFigures}
\end{figure}

Once the concentration of TNF-$\alpha$ increases, glutamate uptake by nearby astrocytes decreases and post synaptic presence increases. This leads to excitotoxicity of the primary neurotransmitter. We have identified a study by Zou et al. \cite{zou2005tnfalpha} which presents some evidence of glutamate uptake reduction due to increase in cytokine concentration. Since there is no explicit data measuring excitotoxicity, we have hypothesized a metric which states that the reduction in uptake is inversely proportional to the increase in extracellular glutamate concentration. Using this principle we have quantified the increase in glutamate as a function of TNF-$\alpha$. The specific mathematical form representing the source term for Eq.\ref{eq:glu} is expressed as: 
\begin{align}
f(c_t) = Cc_t 
\end{align}
where $C = 100 \text{M}^{-1}$ is a fitting parameter. The reaction profile is highlighted in Fig.\ref{fig:allFigures}(D). 

\section{Results}
\label{section:results}
We proceed onto numerically implementing the multiphysics based variational formulation in a standard FE setting in two dimensions. In order to achieve visualisation of the chemical species' evolution firstly we need to geometrically allocate regions within the given domain. These localised domains will signify neurons, microglia and the ECM and will correspond to presence of specific chemical fields in those regions. In order to construct a geometry which resembles a neuron-microglia-ECM assembly at single and multi-neuron length scale we subdivide a given domain into realistic shapes highlighted by the Gaussian point. This approach allow spatio-temporal visualisation of chemical species' diffusion and interaction with the underlying mechanics in a realistic morphological setting, as observed in biological systems. Furthermore, simulation of boundary value problems enables a wide class of kinematics/mechanics driven mechano-chemical phenomena to be analysed.   
\subsubsection*{Single neuron-microglia-ECM morphology}
\begin{figure}[h!]
    \centering
    \includegraphics[width=0.8\linewidth]{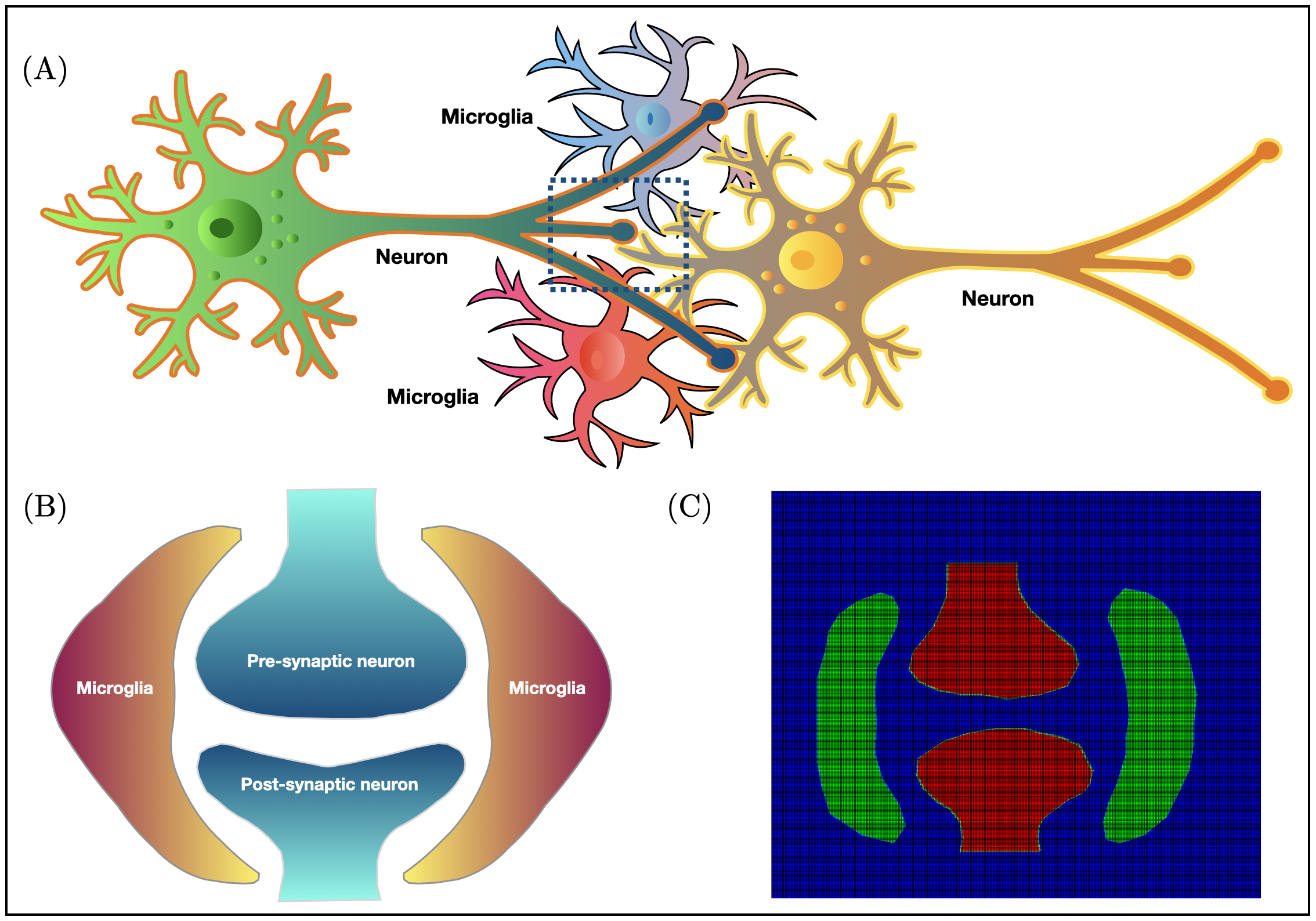}
    \caption{The design of the computational domain has been adapted from a realistic representation of the neurons and the surrounding microglia. As illustrated in \textbf{(A)} a specific region consisting of neuron synapses and microglia has been considered which has been magnified in \textbf{(B)}. A schematic of pre-synaptic and post-synaptic neurons are shown flanged on both sides by microglia. This design has been incorporated as our computational domain \textbf{(C)} to facilitate localisation of chemical fields, visualization of diffusive behaviour of fields and understand interactions between these species at different length and time scales. A meshed version of the numerical domain has been shown with colors (green:microglia, blue:ECM, red:synapse) indicating distinct regions of interest.}
    \label{fig:singleneuron}
\end{figure}

We have illustrated a morphological representation of single neuron-glia-ECM assembly (Fig.\ref{fig:singleneuron}). This consists of neuron synapses indicated by red, padded on both sides by microglia painted in green. The microglia is a pivotal region in numerical simulations, as it facilitates release of ATP to the ECM, houses the P2X7 receptor and modulates channel sensitivity and finally enables production of inflammatory cytokines. All the above mentioned chemical fields are spatially localised in the microglial region and the numerical framework allows diffusion across the ECM. For a single neuron-microglia-ECM assembly we observe the effects of various mechanical loading conditions on this domain through evolution of the underlying chemical and mechanical fields. 
Upon tensile loading, the biomarkers represented by chemical fields diffuse across the illustrated domain. 

\subsubsection*{Multiple neuron-microglia-ECM geometry}
\begin{figure}[h!]
    \centering
    \includegraphics[width=0.8\linewidth]{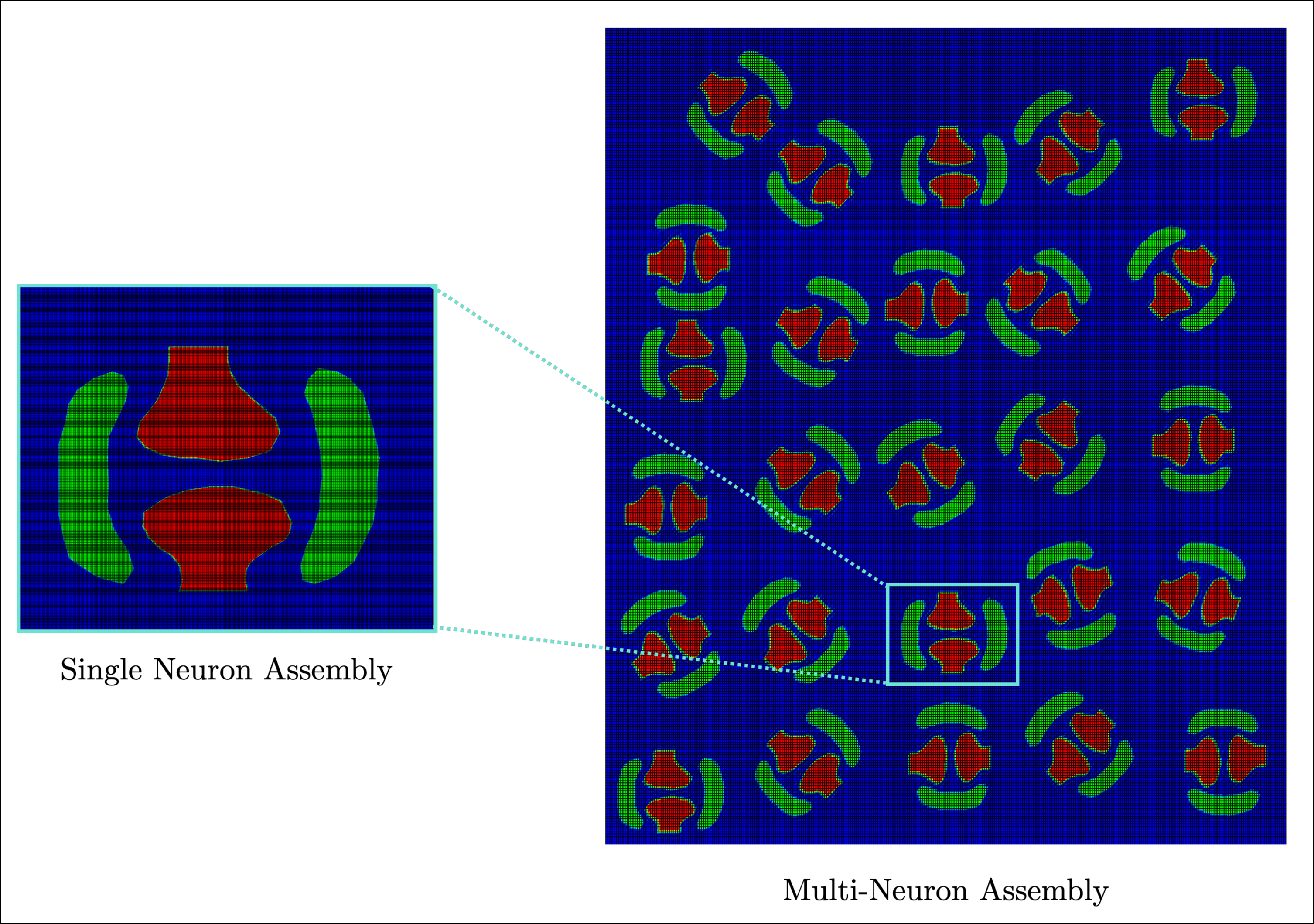}
    \caption{Illustration of a meshed neuron cluster domain (single neuron assembly) in a random spatial distribution of single neuron-microglia assemblies at different orientations (multi-neuron assembly). The dimensions of the single neuron domain are $12 \mu m$ X $10 \mu m$ and that of the larger domain are $60 \mu m$ X $72 \mu m$ }
    \label{fig:singleMultiNeuron}
\end{figure}

In a more realistic setting, the length scale of the domain is increased to accommodate more number of neuron-glia-ECM assemblies which illustrate a better representation of neuron clusters (Fig.\ref{fig:singleMultiNeuron}). As we observe from this construction, more assemblies have been added at random orientations each having the characteristic spatial localisation as mentioned in the single neuron setup. For simplicity we have considered all neuron assemblies of the same type, meaning, the chemical and mechanical properties are the same for all assembly configurations. Primarily, this improvised structure allows us to visualise the spatio-temporal heterogeneity of chemical field evolution and capture essential chemical and mechanical injury metrics across the domain. Similar to the previous setup, we perform different BVPs on this domain and observe field evolution characteristics.   
 
\subsection*{Simulations}
\begin{figure}[hbt!]
    \centering
    \includegraphics[width=\linewidth]{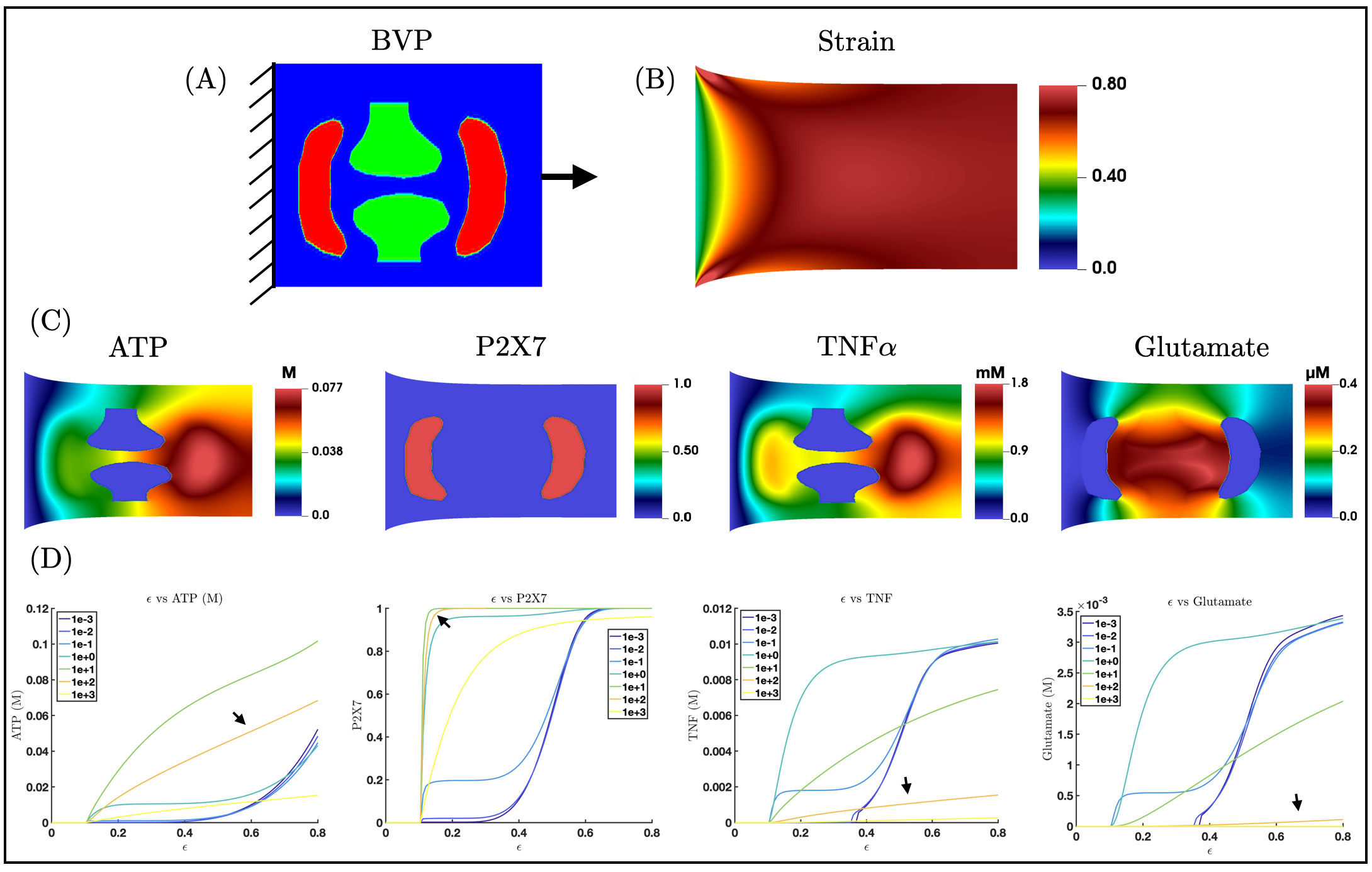}
    \caption{\textbf{(A)} Boundary value problem demonstrating simple tension on a computational domain represented by neuron synapse (green), microglia (red) and ECM (blue). The domain is fixed at one end, while the other end is displaced through a combination of strain and strain-rate loading conditions. \textbf{(B)} Uniaxial strain ($\epsilon_{11}$) field profile obtained from the underlying viscoelastic network by applying displacement through a strain rate of $10^2 s^{-1}$ upto a strain of 0.8. \textbf{(C)} Chemical field evolution profiles representing (left to right) ATP (originating from microglia), P2X7 (localized in microglia), TNF-$\alpha$ (originating from microglia) and Glutamate (originating from neuron synapses). \textbf{(D)} Field evolution plots against a strain of 0.8 has been illustrated for increasing strain rates spanning over six orders of magnitude ($10^{-3}$ - $10^{3}$). The strain rate measure for which the field plots are highlighted is indicated in each subplot and mentioned accordingly.  
}
\label{fig:singleNeuronResults}
\end{figure}

For single and multiple neuron-microglia-ECM assembly, we demonstrate key results based on unaxial tension numerical simulations. For single assembly, as shown in Fig.\ref{fig:singleNeuronResults}(A), the domain is fixed at one end and a displacement-type load is applied across a spectrum of strain and strain-values. In this simulation we highlight results obtained at a strain rate of $10^2 s^{-1}$ upto $80\%$ strain. A fundamental metric of deformation based loading problems is strain; accordingly the axial strain field $(\epsilon_{11})$, obtained from the underlying mechanical network is plotted for the domain as given in Fig.\ref{fig:singleNeuronResults}(B). Using a combination of strain and strain-rates, an array of results are presented in Fig.\ref{fig:singleNeuronResults}(C). The strain field induces ATP to release from the microglia to the ECM. From left to right, field characteristics for ATP due to the above mentioned strain field is shown. As observed, ATP diffuses out of the microglia into the ECM demonstrating the heterogeneous nature of field distribution. The following sub-figure shows P2X7 characteristics which is operating at maximum potential as the critical concentration of ATP, needed to trigger the receptor, has already reached. To its right, TNF-$\alpha$ field plot is shown, which depends on P2X7 receptor opening probability and diffuses out of the microglia. The final sub-plot highlights glutamate concentration due to the influence of TNF-$\alpha$. The neurotransmitter is localised at the synaptic region and diffuses into the ECM. While Fig.\ref{fig:singleNeuronResults}(C) demonstrates full field profiles for each chemical entity, Fig.\ref{fig:singleNeuronResults}(D) illustrates field evolution characteristics for increasing strain-rates from $10^{-3}$ up to $10^{3}$ keeping the maximum strain fixed at 0.8 for each simulation. The following sub-plots (left to right) describe the nature of evolution of ATP (near the microglia), P2X7 (within the microglia), TNF-$\alpha$ (near the microglia) and Glutamate (near the synapse). The particular strain and strain-rate combination used for the full field plots has been highlighted by an arrow in each sub-plot. The maximum magnitude of the chemical species' concentrations are comparable with pathological estimates and presented in Table.\ref{tab:tableTwo}. \par
\begin{figure}[hbt!]
    \centering
    \includegraphics[width=0.7\linewidth]{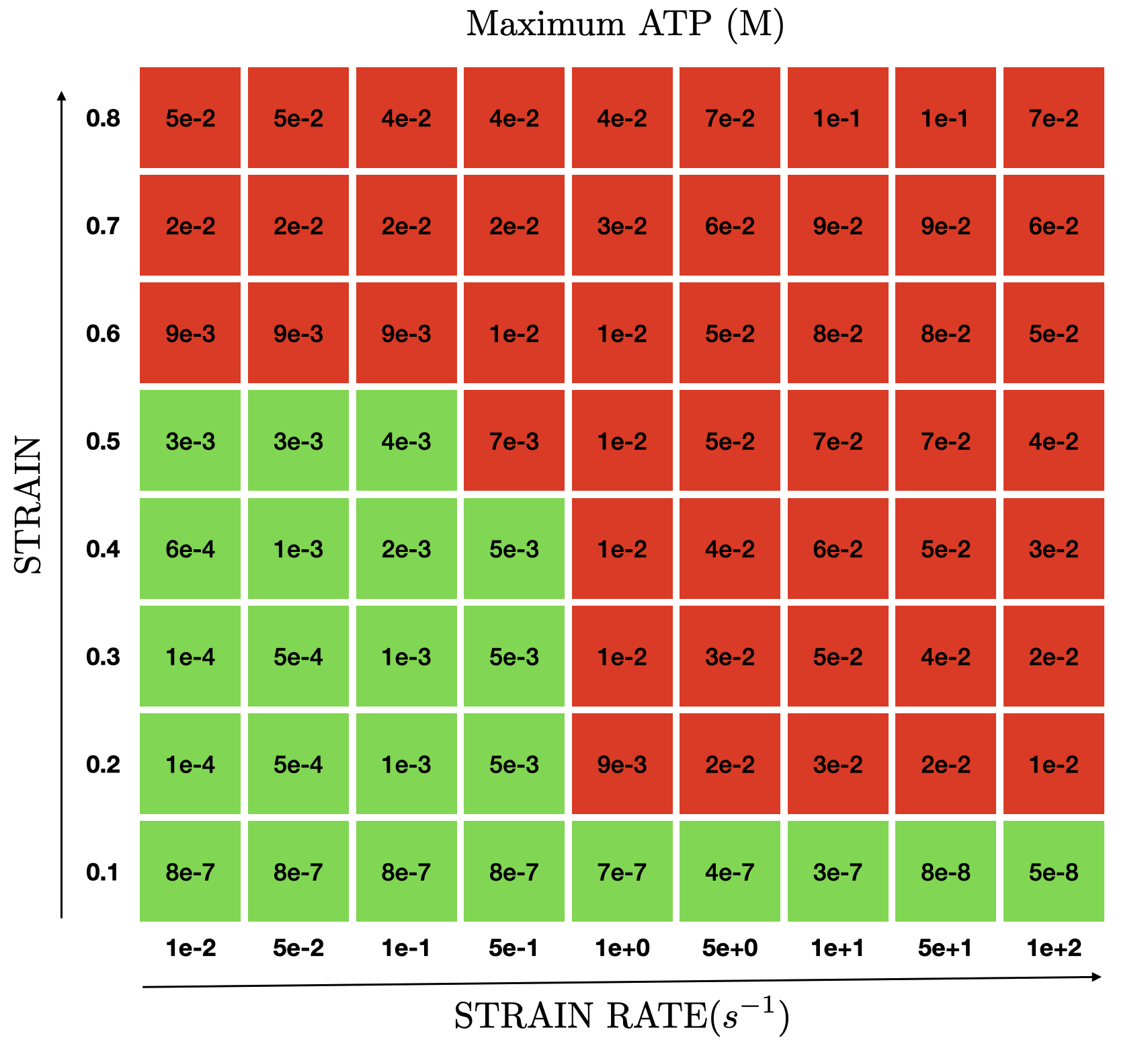}
    \caption{Maximum concentration of ATP within the field distribution is recorded for a spectrum of strain and strain-rates. The resulting data set obtained is utilised to construct a computational injury curve.  Using a specific cut-off for ATP concentration ($6e-3$M) two regions are obtained. The red one represents pathway induced injury while the green one reflects uninjured regimes. }
    \label{fig:injuryCurve}
\end{figure}

A key highlight of the spatio-temporal resolution of these chemical fields is the construction of an injury curve. Considering maximum ATP concentration across the spatial domain at specific strain and strain-rates a dataset is established. The data set is plotted against strain vs strain-rate as illustrated in Fig.\ref{fig:injuryCurve} and allows visual representation of field values at specific loading conditions. A cutoff for critical extracellular ATP is chosen based on the concentration at which purinergic receptors (P2X7) activate \cite{browne2013too}, which is about $6e-3$ mM \cite{xing2016modeling}. The proposition is, if the maximum concentration of ATP stays above this number, the neuronal micro environment is susceptible to chemical degradation, which is also how we define chemical injury. Using this metric, two regions are shown representing pathway induced injury (red boxes) and uninjured region (green boxes). It is a first order estimation of how mechanical loading conditions can influence downstream pathways and elevate critical molecular concentrations. \par
\begin{figure}[th!]
    \centering
    \includegraphics[width=0.7\linewidth]{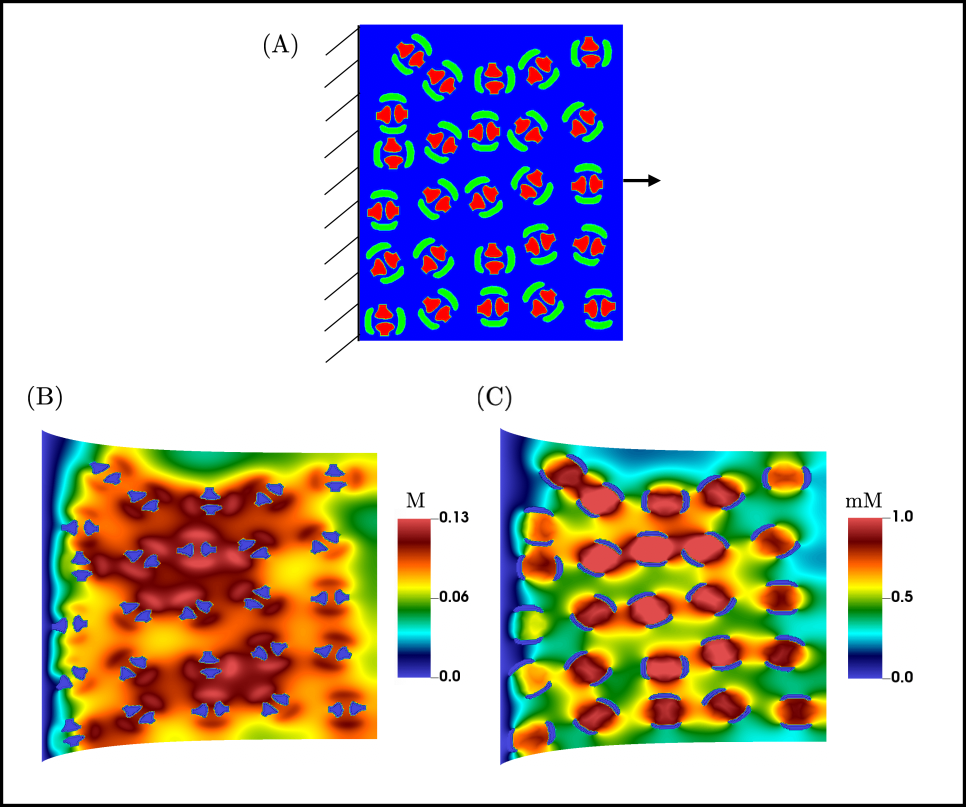}
    \caption{\textbf{(A)} Boundary value problem of multi neuron-glia-ECM assembly is shown. Similar to the single neuron assembly, the larger domain is fixed at one end and displacement based loading is applied at the other end as indicated. \textbf{(B)} ATP evolution for multiple assemblies are shown as induced by the underlying strain field. The heterogeneous nature of field evolution is a hallmark characteristic for such larger domains and can be spatio-temporally resolved to better understand neuron-cluster based experiments. \textbf{(C)} Glutamate evolution is highlighted which is localised near the neuron synapses. Cluster based spatio-temporal resolution of chemical fields such as glutamate serve as key indicators in excitoxicity prediction.}
    \label{fig:multiNeuron}
\end{figure}

Unaxial tensile loading numerical simulations using rate-dependent loading are performed for multi-neuron assemblies. Using variable orientations for individual single neuron-microglia structures, a neuron cluster is constructed and a boundary value problem is setup as illustrated in Fig.\ref{fig:multiNeuron}(A). We have chosen two essential field variables namely ATP (Fig.\ref{fig:multiNeuron}(B)) and glutamate (Fig.\ref{fig:multiNeuron}(C)) to demonstrate the heterogeneity and interactions within this assembly. The results consisting of these field profiles are generated using a strain rate of $10^2 s^{-1}$, applied up to a strain of 0.4.   
\begin{table}[hbt!]
\caption{Comparison of chemical concentrations}
\label{tab:tableTwo} 
\centering
\begin{tabular}{|c|c|c|}
\hline
\textbf{Species} & \textbf{Pathological estimate} & \textbf{Numerical results} \\ \hline  
      ATP   &  1e-2 M \cite{xing2016modeling}   &  7e-2 M        \\ \hline 
      P2X7 &  0 - 1 \cite{xing2016modeling}   &        0 - 1  \\ \hline 
      TNF-$\alpha$   &   2.4 mM \cite{mogi1994tumor}   &  1.8 mM\\ \hline  
   Glutamate  &   1-2 $\mu$M \cite{mark2001pictorial}  &   0.5 $\mu$M      \\ \hline   
\end{tabular}
\end{table}
 
\section{Conclusion}
\label{section:conclusion}
We have proposed and quantified a glia-initiated injury pathway beginning with mechanical deformation and culminating in necroptosis. The proposed model incorporates a multiphysics formulation for investigating mechano-chemical interactions during TBI at the cellular scale. The fidelity of our numerical modeling framework involves a first-of-its-kind representation of neuronal microenvironment as a viscoelastic network coupled with an ensemble of molecular biomarkers represented as chemical fields. Using our modeling framework, we have demonstrated field behaviour for single neuron-glia assembly along with neuron-glia clusters highlighting localisation of biomarkers.  \par 
We believe that the idea of a mechano-chemical  framework utilised to capture pathway induced injury opens up avenues of research directions in injury specific biomarker identification. Although our modeling approach relies on limited availability of biomarker datasets, experimental quantification for injury metrics in human subjects and generic assumptions on field evolution characteristics, it provides a robust numerical and computational base for further improvements. Additionally, our model is capable of incorporating additional biophysics at the neuronal scale including ionic conduction \cite{gulati2023spatio}, nutrient transport and inter-neuronal interactions which possess immense potential in understanding crucial neurodegenerative disorders. 


  
\bibliographystyle{elsarticle-num-names}
\bibliography{ref}
\end{document}